\documentclass[a4paper]{article}
\usepackage{amssymb}

\def\nad#1{\mbox{\smash{\oalign{$#1$\crcr\hidewidth$\mathchar"017E$
\hidewidth}}}}
\def\Nad#1{\mbox{\smash{\oalign{$#1$\crcr\hidewidth$\,\mathchar"707E$
\hidewidth}}}}

\begin{document}

\title{Are neutrinos spinorial tachyons?}
\author{Jakub Rembieli\'nski\thanks{This work is supported under the
{\L}\'od\'z University grant no.\ 457.}\\
Katedra Fizyki Teoretycznej, Uniwersytet {\L}\'odzki\\
ul.~Pomorska 149/153, 90--236 {\L}\'od\'z, Poland\thanks{{\it E-mail
address\/}: jaremb@mvii.uni.lodz.pl}}
\date{}
\maketitle

 \begin{abstract}
 Quantum field theory of space-like particles is investigated in the
framework of absolute causality scheme preserving Lorentz symmetry. It
is shown that tachyons are associated with unitary orbits of Poincar\'e
mappings induced from $SO(2)$ little group instead of $SO(2,1)$ one.
Therefore the corresponding elementary states are labelled by helicity. A
particular case of the helicity $\lambda=\pm\frac{1}{2}$ is investigated in
detail and a corresponding consistent field theory is proposed.
In particular, it is shown that the Dirac-like equation proposed by Chodos
{\it et al}. \cite{CHK}, inconsistent in the standard formulation of QFT, can
be consistently quantized in the presented framework. This allows us to treat
more seriously possibility that neutrinos can be fermionic tachyons as it is
suggested by the present experimental data about neutrino masses \cite{PDG}.
 \end{abstract}

\section{Introduction}
 Almost all recent experiments, measuring directly or indirectly the
electron and muon neutrino masses, have yielded negative values for the
mass square \cite{PDG,CHK}. It suggests that these particles might be
fermionic tachyons. This intriguing possibility was written down some
years ago by Chodos {\em et al}.\ \cite{CHK} and Recami {\em et al}.\
\cite{GMMR}. In the light of the mentioned experimental data we observe
a return of interest in tachyons \cite{Kos,Rem:tac}.

 On the other hand, in the current opinion, there is no satisfactory
theory of superluminal particles. This persuasion creates a
psychological barrier to take such  possibility seriously. Even if we
consider eventuality that neutrinos are tachyons, the next problem
arises; namely a modification of the theory of electro-weak interaction
will be necessary in such a case. But, as we known, in the standard
formulation of special relativity, the unitary representations of the
Poincar\'e group, describing fermionic tachyons, are induced from
infinite dimensional unitary representations of the non-compact $SO(2,1)$
little group.  Consequently, in the conventional approach, the neutrino
field should be infinite-component one so a construction of an
acceptable local interaction is extremaly difficult.

 In this paper we suggest a solution to the above dilemma. To do this we
use the formalism developed in the papers \cite{Rem:tac,Rem:neu} based on the
earlier works \cite{Rem1,Rem2}, where it was proposed a consistent
description of tachyons on both classical and quantum level. The basic
idea is to extend the notion of causality without a serious change of
special relativity. This can be done by means of a freedom in the
determination of the notion of the one-way light velocity, known as the
``conventionality thesis'' \cite{Rei,Jam}.

 In the presented approach the relativity principle is formulated in the
framework of a non-standard synchronization scheme (the
Chang--Tangherlini (CT) scheme). This allows to introduce an absolute
causality for all kinds of events (time-like, light-like, space-like).
For ``standard particles'' our scheme is fully equivalent to the usual
formulation of special relativity. On the other hand, for tachyons it is
possible to formulate covariantly proper initial conditions and there
exists a covariant lower bound of energy. Moreover, the paradox of
``transcendental'' tachyons does not appear in this scheme. On the
quantum level tachyonic field can be consistently quantized using CT
synchronization procedure and they distinguish a preferred frame {\em
via\/} mechanism of the relativity principle breaking \cite{Rem1,Rem:tac},
however with the preservation of the Lorentz covariance and symmetry.

 The main properties of the presented formalism are in the agreement
with {\em local\/} properties of the observed world; namely, we can in
principle distinguish locally a preferred inertial frame by investigation
of the isotropy of the Hubble constant. In fact, it coincides with the
frame in which the Universe appears spherically\footnote{It is well
known such a situation is typical for Robertson--Walker space-times, see
e.g.\ \cite{Wei}.}. Obviously, such a (local) preferred frame should
correlate with the cosmic background radiation frame. Moreover,
present cosmological models incorporate an absolute time (cosmological
time). Therefore it is very natural to look for a local (flat space)
formalism incorporating both Lorentz covariance and a distinguished
inertial frame. Notice that two paradigms of the standard understanding
of the (flat) space-time, namely the assumption of equivalence of
inertial reference frames and a ``democracy'' between time and space
coordinates, are in conflict with the above mentioned local properties of the
observed world.

 In this paper we classify all possible unitary Poincar\'e mappings for
space-like momenta. The important and unexpected result is that unitary
orbits for space-like momenta are induced from the $SO(2)$ little
group. This holds because we have a bundle of Hilbert spaces rather than
a single Hilbert space of states. Therefore unitary operators
representing Poincar\'e group act in irreducible orbits in this bundle.
Each orbit is generated from subspace with $SO(2)$ stability group.
Consequently, elementary states are labelled by helicity, in an analogy
with the light-like case. This fact is extremely important because we
have no problem with infinite component fields.

 Now, let us begin with a brief review of the theory proposed in
\cite{Rem:tac,Rem1,Rem2}.

\section{Formalism}
 According to the papers \cite{Rem:tac,Rem1}, transformation between two
coordinate frames $x^{\mu}$ and ${x'}^\mu$ has the following form
 \begin{equation}\label{1a}
 x'=D(\Lambda,u)(x+a),
 \end{equation}
 \begin{equation}\label{1b}
 u'=D(\Lambda,u)u.
 \end{equation}
 Here $\Lambda$ belongs to the Lorentz group $L$, whilst $u$ is a
four-velocity of a privileged inertial frame\footnote{A necessity of a
presence of a preferred frame for tachyons was stressed by many authors
(see, for example, \cite{CKPG,Sud}).}, as measured by an observer using
$x^{\mu}$ coordinates. The $a^{\mu}$ are translations. The
transformations (\ref{1a}--\ref{1b}) have standard form for rotations i.e.
$D(R,u)=R$, whereas for boosts the matrix $D$ takes the form \cite{Rem:tac}
 \begin{equation}\label{2}
 D(\vec{V},u)=\left(\begin{array}{c|c} \gamma & 0\\[1ex]
\hline \displaystyle-\frac{\vec V}{c}\gamma^{-1} & I+
\displaystyle\frac{\vec V\otimes\vec V^{\rm T}}
{c^2\gamma\left[\gamma+\sqrt{\gamma^2+\displaystyle\frac{\vec V}{c}^2}
\right]}-\frac{\vec V\otimes\vec\sigma^{\rm T}}{c^2\gamma\gamma^2_0}
 \end{array}\right)
 \end{equation}
 where we have used the following notation
 \begin{equation}\label{3a}
 \gamma_0=\left[\frac{1}{2}\left(1+\sqrt{1+\left(\displaystyle
\frac{2\vec\sigma}{c}\right)^2}\right)\right]^{1/2}=\frac{c}{u^0},
 \end{equation}
 \begin{equation}\label{3b}
 \gamma(\vec V)= \left(\left(1+\frac{\vec\sigma\vec V}{c^2}
\gamma^{-2}_0\right)^2 -\left(\frac{\vec
V}{c}\right)^{2}\right)^{1/2},
 \end{equation}
 \begin{equation}\label{3c}
 \frac{\vec{\sigma}}{c}=\frac{\vec{u}}{u^0}.
 \end{equation}
 Here $\vec{V}$ is the relative velocity of $x'$ frame with respect to
$x$ whilst $\vec{\sigma}$ is the velocity of the preferred frame
measured in the frame $x$. The transformations (\ref{2}) remain
unaffected the line element
 \begin{equation}\label{4}
ds^2=g_{\mu\nu}(u)dx^{\mu}dx^{\nu}
 \end{equation}
 with
 \begin{equation}\label{5}
 g(u)=\left(\begin{array}{c|c} 1 & \displaystyle\frac{u^{0}\vec{u}^{\rm
T}}{c^{2}}\\[1ex] \hline \displaystyle\frac{u^{0}\vec{u}}{c^{2}} &
\displaystyle-I+\frac{\vec{u}\otimes\vec{u}^{\rm T}}{c^{4}} (u^{0})^{2}
 \end{array}\right)=\left(\begin{array}{c|c} 1 &
\displaystyle\frac{\vec\sigma^{\rm T}}{c}\gamma_0^{-2}\\[1ex] \hline
\displaystyle\frac{\vec\sigma}{c}\gamma_0^{-2}&
\displaystyle-I+\frac{\vec\sigma\otimes\vec\sigma^{\rm T}}{c^{2}}
\gamma_0^{-4}
 \end{array}\right),
 \end{equation}
 Notice that $u^2=g_{\mu\nu}(u)u^{\mu}u^{\nu}=c^2$.

 From (\ref{4}) we can calculate the velocity of light propagating in a
direction $\vec{n}$
 \begin{equation}\label{6}
 \vec c=\frac{c\vec n}{1-\displaystyle\frac{\vec n\vec\sigma}{c}
\gamma_0^{-2}}.
 \end{equation}
 It is easy to verify that the average value of $\vec{c}$ over a closed
path is always equal to $c$.

 Now, according to our interpretation of the freedom in realization of
the Lorentz group as freedom of the synchronization convention, there
should exists a relationship between $x^{\mu}$ coordinates and the
Einstein-Poincar\'e (EP) ones denoted by $x^{\mu}_{E}$. Indeed, we
observe, that the coordinates
 \begin{equation}\label{7a}
 x_E=T^{-1}(u)x,
 \end{equation}
 \begin{equation}\label{7b}
 u_E=T^{-1}(u)u,
 \end{equation}
 where the matrix $T$ is given by
 \begin{equation}\label{8*}
 T(u)=\left(\begin{array}{c|c}
 1&-\displaystyle\frac{u^{0}\vec{u}^{\rm T}}{c^{2}}\\[1ex]
 \hline
 0&I
 \end{array}\right)=\left(\begin{array}{c|c}
1&\displaystyle-\frac{\vec\sigma^{\rm T}}{c}\gamma_0^{-2}\\[1ex]
\hline
0&I\end{array}\right).
 \end{equation}
 transform under the Lorentz group standardly i.e. (\ref{1a}--\ref{1b}) and
(\ref{7a}--\ref{7b}) imply
 \begin{equation}\label{8a}
x^{\prime}_{E}=\Lambda x_E,
 \end{equation}
 \begin{equation}\label{8b}
 u^{\prime}_{E}=\Lambda u_E.
 \end{equation}
 It holds because $D(\Lambda, u) = T(u') \Lambda T^{-1}(u)$. Moreover,
$ds^2=ds^{2}_{E},\quad \vec{c}_E=c \vec{n},\quad u_{E}^{2}=c^2$ and
$g_E=\eta\equiv{\rm diag}(+,-,-,-)$. Thus the CT synchronization scheme,
defined by the transformations rules (\ref{1a}--\ref{1b}), is at first glance
equivalent to the EP one. In fact, it lies in a different choice of the
convention of the one-way light propagation (see (\ref{6})) under
preserving of the Lorentz symmetry. Notwithstanding, the equivalence is
true only if we exclude superluminal signals. Indeed, the causality
principle, logically independent of the requirement of Lorentz
covariance, is not invariant under change of the synchronization
(\ref{7a}--\ref{7b}). It is evident from the form of the boost matrix
(\ref{2});
the coordinate time $x^0$ is rescaled by a positive fact $\gamma$ only.
Therefore $\varepsilon(dx^0)$  is an invariant of (\ref{1a}--\ref{1b}) and
this fact allows us to introduce an absolute notion of causality,
generalizing the EP causality. Consequently, as was shown in
\cite{Rem:tac}, all inconsistencies of the standard formalism, related to
the superluminal propagation, disappear in this synchronization scheme.

 If we exclude tachyons then, as was mentioned above, physics
cannot depend of synchronization. Thus in this case {\em any
inertial frame can be chosen as the preferred frame \/},
determining a concrete CT synchronization. This statement is in
fact the relativity principle articulated in the CT
synchronization language.

 What happens, when tachyons do exist? In such a case the relativity
principle is obviously broken: If tachyons exist then only one inertial
frame is the {\em true privileged frame}. Therefore, in this case,
the EP synchronization is inadequate to description of reality; we must
choose the synchronization defined by (\ref{1a}--\ref{6}).  Moreover the
relativity principle is evidently broken in this case as well as the
conventionality thesis: The one-way velocity of light becomes ({\em a
priori}) a really measured quantity.

 To formalize the above analysis, in \cite{Rem:tac,Rem2} it was introduced
notion of the synchronization group $L_S$. It connects different
synchronizations of the CT--type and it is isomorphic to the
Lorentz group:
\begin{equation}\label{9a}
x'=T(u')T^{-1}(u)x=D(\Lambda_S,u)T(u)\Lambda_{S}^{-1}T^{-1}(u)x,
\end{equation}
\begin{equation}\label{9b}
u'=D(\Lambda_S,u)u,
\end{equation}
with $\Lambda_S\in L_S$.

 For clarity we write the composition of transformations of the
Poincar\'e group $L\ltimes T^4$ and the synchronization group $L_S$ in
the EP coordinates
\begin{equation}\label{10a}
x_{E}^{\prime}=\Lambda(x_E+a_E),
\end{equation}
\begin{equation}\label{10b}
u_{E}^{\prime}=\Lambda_S\Lambda u_E.
\end{equation}
 Therefore, in a natural way, we can select three subgroups:
 \[
L=\{(I,\Lambda)\},\quad L_S=\{(\Lambda_S,I)\},\quad
 L_0=\{(\Lambda_{0},\Lambda^{-1}_{0})\}.
\]
 By means of (\ref{10a}--\ref{10b}) it is easy to check that $L_0$ and $L_S$
commute.   Therefore the set $\{(\Lambda_S,\Lambda)\}$ is simply the
direct product of  two Lorentz groups $L_0\otimes L_S$.  The
intersystemic Lorentz symmetry group $L$ is  the diagonal subgroup in
this direct product.  From the composition law  (\ref{10a}--\ref{10b})
it follows that $L$ acts as an automorphism group of $L_S$.

 Now, the synchronization group realizes in fact the relativity
principle: If we {\em exclude tachyons} then transformations of
$L_S$ are canonical ones. On the other hand, if we {\em include
tachyons} then the synchronization group $L_S$ is broken to the
$SO(3)_u$ subgroup of $L_S$; here $SO(3)_u$ is the stability group of
$u^{\mu}$. In fact, transformations from the $L_S/SO(3)_u$ do
not leave the absolute notion of causality invariant. On the
quantum level $L_S$ is broken
down to $SO(3)_u$ subgroup i.e. transformations from $L_S/SO(3)_u$
cannot be realized by unitary operators \cite{Rem:tac,Rem2}.

\section{Quantization}
 The following two facts, true only in CT synchronization, are
extremely important for quantization of tachyons \cite{Rem:tac,Rem2}:
 \begin{itemize}
 \item Invariance of the sign of the time component of the space-like
four-mo\-men\-tum i.e. $\varepsilon(k^0)=\mbox{inv}$,
 \item Existence of a covariant lower energy bound; in terms of the
contravariant space-like four-momentum $k^{\mu}$, $k^{2}<0$, this lower
bound is exactly zero, i.e . $k^{0}\geq0$ as in the lime-like and
light-like case.
 \end{itemize}
 This is the reason why an invariant Fock construction can be done in
our case \cite{Rem:tac,Rem2}. In the papers \cite{Rem:tac,Rem2} it was
constructed a quantum free field theory for scalar tachyons. Here we classify
unitary Poincar\'e mappings in the bundle of Hilbert spaces $H_u$ for a
space-like four-momentum. Furthermore we find the corresponding
canonical commutation relations. As result we obtain that tachyons
correspond to unitary mappings which are induced from $SO(2)$ group
rather than $SO(2,1)$ one.  Of course, a classification of unitary
orbits for time-like and light-like four-momentum is standard, i.e., it
is the same as in EP synchronization; this holds because the relativity
principle is working in these cases (synchronization group is unbroken).

\subsection{Tachyonic representations}
 As usually, we assume that a basis in a Hilbert space  $H_u$
(fibre) of one-particle states consists of
the eigenvectors $\left|k,u;\dots\right>$ of the four-momentum
operators\footnote{Notice that we have contravariant as well as
covariant four-momenta related by $g_{\mu\nu}$; the physical
energy and momentum are covariant because they are generators of
translations.} namely
 \begin{equation}
P^{\mu}\left|k,u;\dots\right>=k^{\mu}\left|k,u;\dots\right>        \label{11}
\end{equation}
where
\begin{equation}
\left<k',u;\dots|k,u;\dots\right>=2k_{+}^{0}\delta^{3}(\nad{k'}-\nad{k})
                                                               \label{12}
\end{equation}
 i.e. we adopt a covariant normalization. The
$k_{+}^{0}=g^{0\mu}k^{+}_{\mu}$ is positive and the energy $k^{+}_{0}$
is the corresponding solution of the dispersion relation
 \begin{equation}
k^2\equiv g^{\mu\nu}k_{\mu}k_{\nu}=-\kappa^2.                  \label{13}
\end{equation}
Namely
\begin{equation}
k_{0+}=-\frac{\vec{u}}{u^0}\nad{k}+\omega_{k}\left(\frac{c}{u^0}\right)^2
\label{14}
\end{equation}
with
\begin{equation}
\omega_k=\frac{u^0}{c}\sqrt{\left(\frac{\vec{u}\nad{k}}{c}\right)^2
+\left(|\nad{k}|^2-\kappa^2\right)}.
\label{15}
\end{equation}
Notice that $k_{+}^{0}=\omega_{k}$ and the range of the
covariant momentum $\nad{k}$ is determined by the following inequality
\begin{equation}
|\nad{k}|\geq\kappa\left(1+\left(\left(\frac{c}{u^0}\right)^2-1\right)
\left(\frac{\vec{u}\nad{k}}
{|\vec{u}||\nad{k}|}\right)^2\right)^{-1/2},
\label{16}
\end{equation}
 i.e.\ values of $\nad{k}$ lie outside the oblate spheroid with
half-axes $a=\kappa$ and $b=\kappa\frac{u^0}{c}$. The covariant
normalization in (\ref{12}) is possible because in CT synchronization
the sign of $k^0$ is an invariant (see the form of the matrix $D$ in
the eq. (\ref{2})). Thus we have no problem with an indefinite norm in
$H_u$.

 Now, $ku\equiv k_{\mu}u^{\mu}$ is an additional invariant. Indeed,
because the transformations of $L_S$ are restricted to $SO(3)_u$
subgroup by causality requirement, and $SO(3)_u$ does not change $u$ nor
$k$, our covariance group reduces to the Poincar\'e mappings (realized
in the CT synchrony). Summarizing, irreducible family of unitary
operators $U(\Lambda,a)$ in the bundle of Hilbert spaces $H_u$ acts on
an orbit defined by the following covariant conditions
 \begin{itemize}
 \item $k^2=-\kappa^2$;
 \item $\varepsilon(k^0)=\mbox{inv}$; for
physical representations $k^0>0$ so $\varepsilon(k^0)=1$ which guarantee a
covariant lower bound of energy \cite{Rem:tac,Rem2}.
 \item $q\equiv\frac{uk}{c}=\mbox{inv}$; it is easy to see that $q$
is the energy of tachyon measured in the privileged frame.
 \end{itemize}
 As a consequence there exists an invariant, positive definite measure
 \begin{equation}
d\mu(k,\kappa,q)=d^4\!k\theta(k^0)\delta(k^2+\kappa^2)\delta(q-\frac{uk}{c})
\label{17}
 \end{equation}
 in a Hilbert space of wave packets.

Let us return to the problem of classification of irreducible
unitary mappings $U(\Lambda,a)$:
\[
U(\Lambda,a)\left|k,u;\dots\right>=\left|k',u';\dots\right>;
\]
 here the pair $(k,u)$ is transported along trajectories belonging to an
orbit fixed by the above mentioned invariant conditions. To follow the
familiar Wigner procedure of induction, one should find a stability
group of the double $(k,u)$. To do this, let us transform $(k,u)$ to the
preferred frame by the Lorentz boost $L_{u}^{-1}$. Next, in the
privileged frame, we rotate the spatial part of the four-momentum to the
$z$-axis by an appropriate rotation $R^{-1}_{\vec{n}}$. As a result, we
obtain the pair $(k,u)$ transformed to the pair $(\Nad{k},\Nad{u})$ with
 \begin{equation}
\Nad{k}=\left(\begin{array}{c}
q\\   0\\   0\\   \sqrt{\kappa^2+q^2}
\end{array}\right),\qquad
\Nad{u}=\left(\begin{array}{c}
c\\  0\\  0\\  0
\end{array}\right).                       \label{18}
\end{equation}
 It is easy to see that the stability group of $(\Nad{k},\Nad{u})$ is
the $SO(2)=SO(2,1)\cap SO(3)$ group. Thus tachyonic unitary
representations should be induced from the $SO(2)$ instead of $SO(2,1)$
group! Recall that unitary representations of the $SO(2,1)$ non-compact
group are infinite dimensional (except of the trivial one). As a
consequence, local fields was necessarily infinite component ones
(except of the scalar one). On the other hand, in the CT synchronization
case unitary representations for space-like four-momenta in our bundle
of Hilbert spaces are induced from irreducible, one dimensional
representations of $SO(2)$ in a close analogy with a light-like
four-momentum case. They are labelled by helicity $\lambda$, by $\kappa$
and by $q$ ($\varepsilon(k^0)=\varepsilon(q)$ is determined by $q$; of
course a physical choice is $\varepsilon(q)=1$).

 Now, by means of the familiar Wigner procedure we determine the
Lorentz group action on the base vectors; namely
 \begin{equation}
U(\Lambda)\left|k,u;\kappa,\lambda,q\right>=
e^{i\lambda\varphi(\Lambda,k,u)} \left|k',u';\kappa,\lambda,q\right>
                                             \label{19}
 \end{equation}
where
 \begin{equation}
 e^{i\lambda\varphi(\Lambda,k,u)}=U\left(R^{-1}_{\Omega\vec{n}}
\Omega R_{\vec{n}}\right)                           \label{20}
 \end{equation}
with
 \begin{equation}
 \Omega=L^{-1}_{u'}\Lambda L_u.                \label{21}
 \end{equation}
 Here $k$ and $u$ transform according to the law (\ref{1a}--\ref{1b}).
The rotation $R_{\vec{n}}$ connects $\Nad{k}$ with $D(L^{-1}_{u},u)k$, i.e.
 \begin{equation}
R_{\vec{n}}\Nad{k}=D(L^{-1}_{u},u)k.            \label{22}
\end{equation}
It is easy to check that $R^{-1}_{\Omega\vec{n}}\Omega
R_{\vec{n}}$ is a Wigner-like rotation belonging to the
stability group $SO(2)$ of $(\Nad{k},\Nad{u})$ and
determines the phase $\varphi$. By means of standard topological
arguments $\lambda$ can take integer or half-integer values only
i.e. $\lambda=0, \pm 1/2, \pm 1, \dots.$

 Now, the orthogonality
relation (\ref{12}) reads
\begin{equation}
\left<k',u;\kappa',\lambda',q'|k,u;\kappa,\lambda,q\right>=
2\omega_k\delta^{3}(\nad{k'}-\nad{k})\delta_{\lambda',\lambda}.
                                                               \label{23}
\end{equation}

\subsection{Canonical quantization}
 Following the Fock procedure, we define canonical commutation relations
\begin{equation}\label{24a}
[a_{\lambda}(k_+,u),a_{\tau}(p_+,u)]_{\pm}=
[a_{\lambda}^{\dagger}(k_+,u),a_{\tau}^{\dagger}(p_+,u)]_{\pm}=0,
\end{equation}
\begin{equation}\label{24b}
\mbox{}[a_{\lambda}(k_+,u),a_{\tau}^{\dagger}(p_+,u)]_{\pm}=
2\omega_k\delta(\nad{k}-\nad{p})\delta_{\lambda\tau},
\end{equation}
 where $-$ or $+$ means the commutator or anticommutator and corresponds
to the bosonic ($\lambda$ integer) or fermionic ($\lambda$ half-integer)
case respectively. Furthermore, we introduce a Poincar\'e invariant
vacuum $\left|0\right>$ defined by
 \begin{equation}\label{25}
\left<0|0\right>=1 \qquad\mbox{and}\qquad
a_{\lambda}(k_+,u)\left|0\right>=0.
\end{equation}
 Therefore the one particle states
\begin{equation}
a_{\lambda}^{\dagger}(k_+,u)\left|0\right>                        \label{26}
\end{equation}
are the base vectors belonging to an orbit in our bundle of
Hilbert spaces iff
\begin{equation}\label{27a}
U(\Lambda)a_{\lambda}^{\dagger}(k_+,u)U(\Lambda^{-1})=
e^{i\lambda\varphi(\Lambda,k,u)}
a_{\lambda}^{\dagger}(k_{+}^{\prime},u^{\prime}),
\end{equation}
\begin{equation}\label{27b}
U(\Lambda)a_{\lambda}(k_+,u)U(\Lambda^{-1})=
e^{-i\lambda\varphi(\Lambda,k,u)}a_{\lambda}(k_{+}^{\prime},u^{\prime}),
\end{equation}
and
\begin{equation}\label{28}
[P_{\mu},a_{\lambda}^{\dagger}(k_+,u)]_{-}=
k_{\mu}^{+}\,a_{\lambda}^{\dagger}(k_+,u).
\end{equation}
Notice that
\begin{equation}
P_{\mu}=\int d^4\!k\,\theta(k^0)\,\delta(k^2+\kappa^2)\,k_{\mu}\left(
\sum_{\lambda}a_{\lambda}^{\dagger}(k,u)a_{\lambda}(k,u)\right)
                                                                \label{29}
\end{equation}
 is a solution of (\ref{28}).

 Let us determine the action of the discrete transformations, space and
time inversions, $P$ and $T$ and the charge conjugation $C$ on the
states $|k,u;\kappa,\lambda,q\rangle$.
 \begin{eqnarray}
 P|k,u;\kappa,\lambda,q\rangle
&=&\eta_{s}|k^{\pi},u^{\pi};\kappa,-\lambda,q\rangle,\label{P}\\
 T|k,u;\kappa,\lambda,q\rangle
&=&\eta_{t}|k^{\pi},u^{\pi};\kappa,\lambda,q\rangle,\label{T}\\
 C|k,u;\kappa,\lambda,q\rangle
&=&\eta_{c}|k,u;\kappa,\lambda,q\rangle_{c},\label{C}
 \end{eqnarray}
 where $|\eta_{s}|=|\eta_{t}|=|\eta_{c}|=1$, $k^{\pi}=(k^{0},-\vec{k})$,
$u^{\pi}=(u^{0},-\vec{u})$, the subscript $c$ means the antiparticle
state and $P$, $C$ are unitary, while $T$ is antiunitary.

 Consequently the actions of $P$, $T$ and $C$ in the ring of the field
operators read
 \begin{eqnarray}
 Pa^{\dagger}_{\lambda}(k,u)P^{-1}
&=&\eta_{s}a^{\dagger}_{-\lambda}(k^{\pi},u^{\pi}),\label{Pa}\\
 Ta^{\dagger}_{\lambda}(k,u)T^{-1}
&=&\eta_{t}a^{\dagger}_{\lambda}(k^{\pi},u^{\pi}),\label{Ta}\\
 Ca^{\dagger}_{\lambda}(k,u)C^{-1}
&=&\eta_{c}b^{\dagger}_{\lambda}(k^{\pi},u^{\pi}),\label{Ca}
 \end{eqnarray}
where $b_{\lambda}\equiv a^c_{\lambda}$---antiparticle operators.

 Finally we can deduce also the form of the helicity operator:
\begin{equation}
 \hat{\lambda}(u)=-\frac{W^{\mu}u_{\mu}} {\displaystyle
c\sqrt{\left(\frac{Pu}{c}\right)^2 -P^2}}
  \label{30}
 \end{equation}
where
\[
W^{\mu}=\frac{1}{2}\varepsilon^{\mu\sigma\lambda\tau}J_{\sigma\lambda}
P_{\tau}
\]
is the Pauli-Lubanski four-vector.

 Notice that
 \begin{eqnarray}
 P\hat{\lambda}(u)P^{-1}&=&-\hat{\lambda}(u^{\pi}),\label{Pl}\\
 T\hat{\lambda}(u)T^{-1}&=&\hat{\lambda}(u^{\pi}),\label{Tl}\\
 C\hat{\lambda}(u)C^{-1}&=&\hat{\lambda}(u),\label{Cl}
 \end{eqnarray}
 as well as
 \begin{equation}
 [\hat\lambda(u),a^\dagger_{\lambda}(u,k)]=\lambda a^\dagger_{\lambda}(u,k).
\label{la}
 \end{equation}

\subsection{Local fields}
 As usually we define local tachyonic fields as covariant Fourier
transforms of the creation--annihilation operators. Namely
 \begin{eqnarray}
 \lefteqn{\varphi_{\alpha}(x,u)=\frac{1}{(2\pi)^{\frac{3}{2}}}
\int_{0}^{\infty}dq\,\rho(q)\int d\mu(k,\kappa,q)\sum_{\lambda}}
\nonumber\\
&\times&\left[w_{\alpha\lambda}(k,u)e^{ikx}b_{\lambda}^{\dagger}(k,u)
+v_{\alpha\lambda}(k,u)e^{-ikx}a_{\lambda}(k,u)\right],\label{field}
 \end{eqnarray}
 where the amplitudes $w_{\alpha\lambda}$ and $v_{\alpha\lambda}$
satisfy the set of corresponding consistency conditions (the Weinberg
conditions). Here we sum over selected helicities and over the invariant $q$
with a measure $\rho(q)dq$. It can be shown that the density $\rho(q)$
determines the form of translation generators $P_{\mu}$ deduced from the
corresponding Lagrangian. On the other hand
$P_{\mu}$ are given by eq.\ (\ref{29}). Both definitions coincide only
for $\rho(q)=1$. The above statement can be easily verified for the
scalar tachyon field discussed in \cite{Rem:tac} and for the fermionic
tachyon field discussed below. Therefore in the following we choose
simply $\rho(q)=1$. Thus the integration in (\ref{field}) reduces to the
integration with the measure
$d^{4}k\,\theta(k^{0})\delta(k^{2}+\kappa^{2})$.

\section{Fermionic tachyons with helicity $\lambda=\pm\frac{1}{2}$}
 To construct tachyonic field theory describing field excitations with
the helicity $\pm\frac{1}{2}$, we assume that our field transforms under
Poincar\'e group like bispinor (for discussion of transformation rules
for local fields in the CT synchronization see \cite{Rem1}); namely
 \begin{equation}
 \psi'(x',u')=S(\Lambda^{-1})\psi(x,u),
 \end{equation}
 where $S(\Lambda)$ belongs to the representation
$D^{\frac{1}{2}0}\oplus D^{0\frac{1}{2}}$ of the Lorentz group. Because
we are working in the CT synchronization, it is convenient to introduce
an appropriate (CT-covariant) base in the algebra of Dirac matrices as
 \begin{equation}
 \gamma^{\mu}={T(u)^\mu}_{\nu}\gamma^{\nu}_{E},
 \end{equation}
 where $\gamma^{\mu}_{E}$ are standard $\gamma$-matrices, while $T(u)$
is given by the eq.\ (\ref{8*}). Therefore
 \begin{equation}
 \{\gamma^{\mu},\gamma^{\nu}\}=2g^{\mu\nu}(u)I.
 \end{equation}
 However, notice that the Dirac conjugate bispinor
$\bar\psi=\psi^\dagger\gamma^0_E$. Furthermore
$\gamma^5=-\frac{i}{4!}\epsilon_{\mu\nu\sigma\lambda}
\gamma^\mu\gamma^\nu\gamma^\sigma\gamma^\lambda=\gamma^5_E$.

 Now, we look for covariant field equations which are of degree one\footnote{In
the Ref.\ \cite{Rem:neu} we found a class of the second order equations under
condition of the $P$-invariance.}
with respect to the derivatives $\partial_{\mu}$ and imply the
Klein--Gordon equation
 \begin{equation}
 \left(g^{\mu\nu}(u)\partial_{\mu}\partial_{\nu}-\kappa^2\right)\psi=0,
 \end{equation}
 related to the space-like dispersion relation $k^{2}=-\kappa^{2}$. We also
requiure the $T$-invariance of these equations.

 As the result we obtain the following family of the Dirac-like equations
 \begin{eqnarray}
\lefteqn{ \left\{\left(\frac{u\gamma}{c}\sin\alpha-1\right)
\left(\left(i\frac{u}{c}\partial\right)\cos\beta
-\kappa\sin\beta\right)\right.}\nonumber\\
\lefteqn{\null\quad\left.-\gamma^5\left[(-i\gamma\partial)
+\frac{i}{2}\left[\gamma\partial,\frac{u\gamma}{c}\right]\sin\alpha
\right.\right.}\nonumber\\
&&\left.\left.+\frac{u\gamma}{c}\left(\left(i\frac{u}{c}\partial\right)
(1+\cos\alpha\sin\beta)+\kappa\cos\alpha\cos\beta\right)\right]\right\}
\psi(x,u)=0, \label{*}
 \end{eqnarray}
 derivable from an appriopriate hermitian Lagrangian density.
 Here $u\gamma=u_\mu\gamma^\mu$, $u\partial=u^\mu\partial_\mu$,
$\gamma\partial=\gamma^\mu\partial_\mu$ and $\alpha$, $\beta$---real
parameters, $\alpha\neq(2n+1)\frac{\pi}{2}$. To guarantee the irreducibility of
the elementary system described by (\ref{*}), the equation (\ref{*}) must be
accompaniated by the {\em covariant\/} helicity condition
 \begin{equation}
 \hat\lambda(u)\psi(u,k)=\lambda\psi(u,k)\label{**}
 \end{equation}
 where $\hat\lambda$ is given by (\ref{30}) taken in the coordinate
representation (see below) and $\lambda$ is fixed ($\lambda=\frac{1}{2}$ or
$-\frac{1}{2}$ in our case). This condition is quite analogous to the
condition for the left (right) bispinor in the Weyl's theory of the massless
field. It implies that particles described by $\psi$ have helicity $-\lambda$,
while antiparticles have helicity $\lambda$. For the obvious reason in the
following we will concentrate on the case $\lambda=\frac{1}{2}$.

 Notice that the pair of equations (\ref{*},\ref{**}) is not invariant under
the $P$ or $C$ inversions separately for every choice of $\alpha$ and $\beta$.

 Now, in the bispinor realization the helicity operator $\hat\lambda$ has the
following explicit form
 \begin{equation}
 \hat\lambda(u)=\frac{\gamma^5\left[-i\gamma\partial,\frac{u\gamma}{c}\right]}
{4\sqrt{\left(-i\frac{u\partial}{c}\right)^2+\square}} \label{***}
 \end{equation}
 where the integral operator
$\left(\left(-i\frac{u\partial}{c}\right)^2+\square\right)^{-\frac{1}{2}}$ in
the coordinate representation is given by the well behaving distribution
 \begin{equation}
 \frac{1}{\sqrt{\left(-i\frac{u\partial}{c}\right)^2+\square}}
=\frac{1}{(2\pi)^4}\int\frac{d^4p\,\varepsilon\left(\frac{up}{c}\right)e^{ipx}}
{\sqrt{\left(\frac{up}{c}\right)^2-p^2}}. \label{****}
 \end{equation}

 Now, let us notice that the equation (\ref{*}), supplemented by the helicity
condition (\ref{**}), are noninvariant under the composition of the $P$ and $C$
inversions (see eqs.\ (\ref{Pa}--\ref{Ca}) and the Appendix), except of the
case $\sin\alpha=\cos\beta=0$. Because (\ref{*}--\ref{**}) are $T$-invariant,
therefore only for $\sin\alpha=\cos\beta=0$ they are $CPT$-invariant. Taking
$\sin\beta=\cos\alpha=1$ we obtain from (\ref{*})
 \begin{equation}\label{!}
 \left\{\kappa+\gamma^5\left[i\gamma\partial
-2\frac{u\gamma}{c}\left(i\frac{u}{c}\partial\right)\right]\right\}\psi=0,
 \end{equation}
 supplemented by (\ref{**}). On the other hand, for $\cos\alpha=-\sin\beta=1$
we obtain
 \begin{equation}\label{!!}
 \left(\kappa-\gamma^5(i\gamma\partial)\right)\psi=0.
 \end{equation}

 The last equation is exactly the Chodos {\it et al}.\ \cite{CHK} Dirac-like
equation for tachyonic fermion. However, contrary to the standard EP approach,
it can be consistently quantized in our scheme (if it is supplemented by the
helicity condition (\ref{**})). In the following we will analyze the eqs.\
(\ref{!!}) and (\ref{**}) by means of the Fourier decomposition
 \begin{equation}
 \psi(x,u)=\frac{1}{(2\pi)^{\frac{3}{2}}}\int d^4k\,\delta(k^2+\kappa^2)
\theta(k^0)\left[w_{\frac{1}{2}}(k)e^{ikx}b^\dagger_{\frac{1}{2}}(k)
+v_{-\frac{1}{2}}(k)e^{-ikx}a_{-\frac{1}{2}}(k)\right] \label{*****}
 \end{equation}
 of the field $\psi$. The creation and annihilation operators $a$ and $b$
satisfy the corresponding canonical anticommutation relations
(\ref{24a}--\ref{24b}), i.e., the nonzero ones are
\begin{equation}\label{a1}
[a_{-\frac{1}{2}}(k),a^\dagger_{-\frac{1}{2}}(p)]_+
=2\omega_k\delta(\nad{k}-\nad{p})
\end{equation}
\begin{equation}\label{a2}
[b_{\frac{1}{2}}(k),b^\dagger_{\frac{1}{2}}(p)]_+
=2\omega_k\delta(\nad{k}-\nad{p})
\end{equation}
In (\ref{*****}) $b_{-\frac{1}{2}}$ and
$a_{\frac{1}{2}}$
do not appear because we decided to fix $\lambda=\frac{1}{2}$ in (\ref{**})
(compare with (\ref{la})). As the consequence of (\ref{**}) the corresponding
amplitudes $w_{-\frac{1}{2}}$ and $v_{\frac{1}{2}}$ vanish. The
nonvanishing amplitudes $w_{\frac{1}{2}}$ and $v_{-\frac{1}{2}}$ satisfy
\begin{equation}\label{E1}
(\kappa+\gamma^5k\gamma)w_{\frac{1}{2}}(k,u)=0,
\end{equation}
\begin{equation}\label{E2}
\left(1-\frac{\gamma^5\left[k\gamma,\frac{u\gamma}{c}\right]}{2
\sqrt{q^2+\kappa^2}}\right)w_{\frac{1}{2}}(k,u)=0,
\end{equation}
\begin{equation}\label{E3}
(\kappa-\gamma^5k\gamma)v_{-\frac{1}{2}}(k,u)=0,
\end{equation}
\begin{equation}\label{E4}
\left(1-\frac{\gamma^5\left[k\gamma,\frac{u\gamma}{c}\right]}{2
\sqrt{q^2+\kappa^2}}\right)v_{-\frac{1}{2}}(k,u)=0.
\end{equation}
Here $k\equiv k_+$, $q=\frac{uk_+}{c}$. The solution of (\ref{E1}--\ref{E4})
reads
\begin{equation}\label{S1}
w_{\frac{1}{2}}(k,u)=\left(\frac{\kappa-\gamma^5k\gamma}{2\kappa}\right)
\frac{1}{2}\left(1+\frac{\gamma^5\left[k\gamma,\frac{u\gamma}{c}\right]}{2
\sqrt{q^2+\kappa^2}}\right)w_{\frac{1}{2}}(\Nad k,\Nad u),
\end{equation}
\begin{equation}\label{S2}
v_{-\frac{1}{2}}(k,u)=\left(\frac{\kappa+\gamma^5k\gamma}{2\kappa}\right)
\frac{1}{2}\left(1+\frac{\gamma^5\left[k\gamma,\frac{u\gamma}{c}\right]}{2
\sqrt{q^2+\kappa^2}}\right)v_{-\frac{1}{2}}(\Nad k,\Nad u),
\end{equation}
where the amplitudes are normalized by the covariant conditions
\begin{equation}\label{N1}
\bar w_{\frac{1}{2}}(k,u)\frac{u\gamma}{c}\gamma^5w_{\frac{1}{2}}(k,u)
=\bar v_{-\frac{1}{2}}(k,u)\frac{u\gamma}{c}\gamma^5v_{-\frac{1}{2}}(k,u)
=2q,
\end{equation}
\begin{equation}\label{N2}
\bar w_{\frac{1}{2}}(k^\pi,u)\frac{u\gamma}{c}\gamma^5v_{-\frac{1}{2}}(k,u)=0.
\end{equation}
The amplitudes $w_{\frac{1}{2}}(\Nad{k},\Nad{u})$ and
$v_{-\frac{1}{2}}(\Nad{k},\Nad{u})$, taken for the values $\Nad k$ and
$\Nad u$ given in the eq.\ (\ref{18}), have the following explicit form (for
$\gamma^\mu_E$ matrix convention---see Appendix)
\begin{equation}\label{form}
w_{\frac{1}{2}}(\Nad k,\Nad u)=\pmatrix{\sqrt{q+\sqrt{q^2+\kappa^2}}\cr
0\cr\frac{-\kappa}{\sqrt{q+\sqrt{q^2+\kappa^2}}}\cr0},\quad
v_{-\frac{1}{2}}(\Nad k,\Nad u)=\pmatrix{\sqrt{q+\sqrt{q^2+\kappa^2}}\cr
0\cr\frac{\kappa}{\sqrt{q+\sqrt{q^2+\kappa^2}}}\cr0}.
\end{equation}
It is easy to see that in the masseless limit $\kappa\to0$ the eqs.\
(\ref{E1}--\ref{E4}) give the Weyl equations
\[ k\gamma w_{\frac{1}{2}}=k\gamma v_{-\frac{1}{2}}=0,\qquad
\gamma^5w_{\frac{1}{2}}=w_{\frac{1}{2}},\quad
\gamma^5v_{-\frac{1}{2}}=v_{-\frac{1}{2}}, \]
as well as the amplitudes (\ref{S1}--\ref{S2}) have a smooth $\kappa\to0$
limit (it is enough to verify (\ref{form})).

Now, the normalization conditions (\ref{N1}--\ref{N2}) generate the proper work
of the canonical formalism. In particular, starting from the Lagrangian density
${\cal L}=\bar\psi\left(\kappa-\gamma^5(i\gamma\partial)\right)\psi$ we can
derive the translation generators; with help of (\ref{*****},\ref{a1},\ref{a2})
and (\ref{N1},\ref{N2}) we obtain
\begin{equation}\label{Pm}
P_\mu=\int\frac{d^3\nad{k}}{2\omega_k}k_\mu(
a^\dagger_{-\frac{1}{2}}a_{-\frac{1}{2}}+
b^\dagger_{\frac{1}{2}}b_{\frac{1}{2}})
\end{equation}
In agreement with (\ref{29}). Thus we have constructed fully consistent free
field theory for a fermionic tachyon with helicity $\pm\frac{1}{2}$, quite
analogous to the Weyl's theory for a left spinor which is obtained as the
$\kappa\to0$ limit.

\section{Conclusions}
 The main result of this work is that tachyons are classified according
to the unitary representations of $SO(2)$ rather than $SO(2,1)$ group;
so they are labelled by the eigenvectors of the helicity operator. In
particular for the helicity $\lambda=\pm\frac{1}{2}$ we have constructed family
of $T$-invariant equations (\ref{*}). Under condition of $PCT$ invariance we
selected two equations (\ref{!}) and (\ref{!!}). The equation (\ref{!!})
coincide with the one proposed by Chodos {\it et al}. \cite{CHK}. We show by
explicit construction that, in our scheme, theory described by this equation,
supplemented by the helicty condition (\ref{**}) can be consistently
quantized. This theory describe fermionic tachyon with helicity $-\frac{1}{2}$.
It has a smooth massless limit to the Weyl's left-handed spinor theory. These
results show that there are no theoretical obstructions to interpret the
experimental data about square of mass of neutrinos \cite{PDG} as a signal that
they can be fermionic tachyons.

\appendix
\section{Appendix}
 The discrete transformations $P$, $T$ and $C$, defined by the eqs.\
(\ref{P}--\ref{C}) are realised in the bispinor space standardly, i.e. $P$ by
$\gamma^0_E$, while $T$ and $C$ by $\cal T$ and $\cal C$ satisfying the
conditions
 \begin{equation}\label{A4}
 {\cal T}^\dagger{\cal T}=I,\quad {\cal T}^*{\cal T}=-I,\quad
{\cal T}^{-1}{\gamma^\mu}^{\rm T}{\cal T}=\gamma^\mu,
 \end{equation}
 \begin{equation}\label{A5}
 {\cal C}^\dagger{\cal C}=I,\quad {\cal C}^*{\cal C}=-I, \quad
{\cal C}^{\rm T}=-{\cal C},\quad
{\cal C}^{-1}\gamma^\mu{\cal C}=-{\gamma^\mu}^{\rm T}.
 \end{equation}
 Notice that the last condition in (\ref{A4}) and (\ref{A5}) can be fortmulated
in terms of  the standard $\gamma^\mu_E$ exactly in the same form.

In explicit calculations of the amplitudes (\ref{form}) we have used the
following representations of the $\gamma_E$ matrices: $\vec\gamma_E=\pmatrix{
0&-\vec\sigma\cr\vec\sigma&0}$, $\gamma^0_E=\pmatrix{0&I\cr I&0}$. In this
representation the parity, charge conjugation and time inversion are
given, up to a phase factor by
\[ {\cal P}=\gamma^0_E,\quad {\cal C}=i\gamma^0_E\gamma^2_E,\quad
{\cal T}=-i\gamma^0_E\gamma^2_E\gamma^5_E.\]

 \bibliographystyle{prsty}
% \bibliography{tachyon}

 \end{document}